# Signal Processing with Pulse Trains: An Algebraic Approach- Part I

Gabriel Nallathambi and Jose C. Principe

*Abstract*— **Recently we have shown that it is possible to represent continuous amplitude, continuous time, band limited signals with an error as small as desired using pulse trains via the integrate and fire converter (IFC). The IFC is an ultra low power converter and processing with pulse trains is compatible with the trends in the silicon technology for very low supply voltages. This paper presents the definition of addition in pulse trains created by the IFC using exclusively timing information, and proofs that it constitutes an Abelian group in the space of IFC pulse trains. We also show that pulse domain addition corresponds to pointwise addition of analog signals.**

*Index Terms*— Algebraic signal processing, integrate and fire sampler, inter-pulse interval, pulse train processing.

## I. Introduction

The dependence of digital representations in signal processing is overwhelming. The reason it is difficult to think of computation without numbers is due to the ubiquitous role of the digital computer, which implements the mathematical operators defined in the Church-Turing computability theory. The Whittaker-Shannon-Nyquist theorem opens the possibility of representing continuous amplitude time signals with finite bandwidth as strings of numbers and allowed the use of digital computers to process real world data. However, the concept of computation is much broader and our brain is a living proof that it can compute without using number based representations (although it can learn how to do arithmetic with numbers during childhood) [1]. By converting continuous time signal to numbers and operating on them with digital computers, our technology has grown drastically in precision and sophistication. This is the well-accepted way of doing computation and it will remain so for many years to come in computer clusters and mainframes because several years of algorithmic development have been invested in this technology. However, growing application domains such as portable computing systems are experiencing constant design tradeoffs as the flexibility and multi-purpose functionality is based on increasingly sophisticated software, which cost clock-cycles. Since clock-cycles cost power, the battery power budget is a limiting factor. Video processing is going to be the next big number crunching application for cellphones, and it will shorten battery life even further.

The other worrisome trend is the slowdown of innovation in silicon technology, the famous Moore's law. We cannot expect to continue seeing a doubling of computer power every 2 years. Moreover, we can expect the dynamic range of silicon devices to decrease to supply voltages of 0.3V~0.4 V or lower and with much higher speeds. This in fact follows the CMOS technology trend: as we move towards scaled nanometer CMOS nodes, the voltage headroom is reduced whereas the time resolution (gate delay) is improved. Based on a recent comprehensive study [2], A/D performance in terms of dynamic range and signal-to-noise ratio (SNR) does not necessarily benefit from these scaled nanometer nodes. All these concerns are being dealt with piecemeal approaches and have not triggered yet a massive reevaluation of other forms of computation.

Von Neumann, who invented the first digital computer, proposed an alternative to binary arithmetic in a 1952 lecture entitled "Probability logic and the synthesis of reliable organisms from unreliable components"[3]. He proved that reliable automata/networks, i.e., networks with a probability of output error $P_e < 0.5$ can be designed using a cascade of three-input majority gates, if the component probability of failure $p_e \leq 0.0073$, and that reliable computation is impossible if $p_e \geq 1/6$. Gaines in 1967 [4], [5] proposed to use the probability of a binary value to encode an analog value and he created the term stochastic computing. A quantity is represented by a clocked sequence of logic levels generated by a random process (Bernoulli sequence). He proposed 3 linear mappings to transform analog variables into probabilities with digital gates (i.e. 0 for zero probability, 1 for maximum range, and the probability in between). Finite state machines could operate with these probabilities to do computation [4], or implementations based on neural networks [6], but one of the problems is poor scaling with precision. More recently, computer scientists interested in low power implementations picked up the idea [7], and a recent survey appears in [8]. We are not aware of other alternatives within the realm of computation theory. Neuromorphic architectures such as the

G. Nallathambi is with the Department of Electrical and Computer Engineering, Gainesville, Florida, 32611, USA (e-mail: gabriel_n,@ymail.com).

J. C. Principe is with the Department of Electrical and Computer Engineering, Gainesville, Florida, 32611, USA (e-mail: principe@cnel.ufl.edu).

Liquid State Machine [9] are silicon implementations of brain style computation. They have been proved universal computers [9], but they use learned mappings and they impose a style of approximations that constrain the type of problems that can be solved, i.e. arithmetic is not one of their strengths, so we will not review this literature.

Our work attempts to present an alternative to digital signal processing by converting analog signal voltages into time between pulses (+/-1), implemented by an integrate and fire converter (IFC) [10], [11] of Fig. 1. First, we proved that it is possible to reconstruct analog signals with finite bandwidth from pulse trains with an error as small as required, mimicking the Nyquist theorem result [10]. This provides a solid foundation to develop a signal processing theory based on pulse-based representations, which is the goal of our work. Therefore a continuous time pulse train can represent a continuous amplitude signal, and the goal is to operate algebraically with these pulse trains. Only timid attempts are known in this direction. During the 1980's, efforts were made to develop pulse train signaling methods to build efficient analog silicon networks. Murray [12], [13] reviewed pulse encoding methods such as pulse amplitude, width, delay and frequency modulation with pulse arithmetic performed at the neuron/synapse level. Miura et al. [14] proposed using magnetic cores with storing and voltage-time integrating property for performing pulse arithmetic. These are outside the scope of this paper as the pulse trains here are generated by an IFC and the arithmetic is performed deterministically in the pulse domain without integrating back to the amplitude domain. Inspired by biology, Lazar group [15]–[17] has worked with sets of IFC with different parameters to create population encoding for single or multiple sensors, and the results are very exciting for video processing, and mimic what is known in the visual cortex. However, his approach is not signal processing (arithmetic) oriented. McCormick [18] has modeled the IFC as a cascade of integrator, uniform quantizer and differentiator. In their formulation to perform arithmetic computations in the pulse domain, a quantized version of the integral has to be reconstructed, and addition/multiplication can be done over the quantized waveform, which yields another quantized waveform. The resulting quantized waveform is converted back to pulses using the differentiator. Unfortunately the pulses created by this differentiated waveform may violate the constraint that the time between pulses is a constant defined by the threshold, i.e. it cannot be produced by an IFC. We propose a radically different, systematic, approach of performing computations in the pulse domain without reconstructing the quantized waveform. This paper discusses about pulse domain addition, and multiplication will be presented in a companion paper.

The rest of the paper is organized as follows: Section II describes the IFC in detail and contrasts it with Nyquist samplers. In section III, we present theoretical foundations for algebraic operations in pulse trains and propose theorems for performing addition in the pulse domain. In section IV, we study various group axioms and show that pulse domain addition constitutes an Abelian group in the space of IFC pulse trains. In section V, we numerically evaluate the theorems and study the behavior of pulse domain addition under variation of different parameters.

## II. INTEGRATE AND FIRE SAMPLER

We have designed a special hardware device called the integrate-and fire converter (IFC) that mimics the way the neuron works, but we assume that its *output is a deterministic signal*. Our goal is to represent the continuous amplitude signal in a compressed manner and create an injective mapping (one to one with a unique inverse) between the two representations, imposing as a constraint finite bandwidth as employed in Nyquist theory. The IFC is a very simple device because it integrates the input and when the voltage at the capacitor reaches a threshold it creates a pulse, and resets the capacitor [19]. For practical purposes we found out that the number of pulses is substantially decreased if the IFC has a positive and negative threshold [20]. The IFC model is presented in Fig. 1

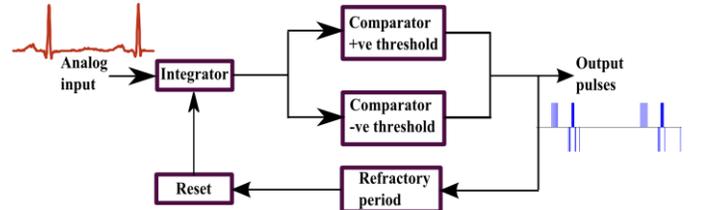

Fig. 1. Block diagram of integrate and fire sampler

The continuous function $f(t)$ is convolved with an "averaging function", $h_k(t)$. The result is compared against two fixed thresholds and when either of these is reached, a pulse is created at time $t_k$ preserving the threshold polarity (positive or negative). The value of the integrator is reset and held at this state depending on the refractory period specified by $\tau$; the process then repeats. The output pulse train is defined over a discrete set of non-uniformly spaced times. Each of the pulse intervals $t_k$ satisfies the condition

$$\theta_{+,-} = \int_{t_{k-1}}^{t_k} f(\tau) h_k(t-\tau) d\tau \qquad (1)$$

One common form of $h_k(t)$ is a first order lowpass filter $\frac{1}{C} e^{\frac{-t}{RC}} u(t_k - t_{k-1})$. The constants $R$ and $C$ represent the resistance and the capacitor values in the hardware implementation of the leaky integrator. To reconstruct the original signal from the samples a basis $\Phi$ is selected in a frame (e.g. splines) and express the approximation as $\hat{f}(t) = \sum_{k=1}^{M} a_k \phi_k(t)$ where $a_k$ are obtained by linear regression $\vec{\theta} = S\vec{a}$, where S is a matrix of terms constructed by integrating the basis set over the time intervals of sequential IFC pulses [10]. We show in [10] that

$$\|f(t) - \hat{f}(t)\|_\infty \leq C\theta \qquad (2)$$

where $C$ is a constant only dependent on the window of analysis and basis functions, and θ is the IFC threshold. We have studied the use of this device as an ADC replacement and found out that the area and power consumption is smaller than most of the ADCs available: a single channel IF has ~ 30 transistors, with a layout box of 100 μm X 100 μm using CMOS 0.6 μm technology and with a figure of merit (FOM) of 0.6 pJ/conv for an 8 bit converter [11]. The pulse train created by the IFC takes advantage of the signal time structure, and at least for the impulsive class of signals it can reduce the sampling rates, for a given reconstruction fidelity, with respect to Nyquist samples [21]–[23]. It can therefore be interpreted as a compressed representation of the analog input signal for this signal class.

It is instructive to compare the IFC with Nyquist samplers. Any finite bandwidth real signal $f(t)$ defined in the real line, can be exactly recovered from the samples $f(nT)$ obtained using delta functions as

$$f(t)|_{t=nT} = \sum_{n=-\infty}^{\infty} \int_{-\infty}^{\infty} f(\lambda)\delta(\lambda - nT)d\lambda \quad (3)$$

when $1/T > 2f_{max}$ [24]. In practice, instead of the delta function, A/D converters use very short pulses of duration $\Delta t$, i.e.

$$f(t)|_{t=nT} \approx \sum_{n=-\infty}^{\infty} \int_{nT}^{nT+\Delta t} f(\lambda)d\lambda \quad (4)$$

Suppose that we constrain the area of the integral to a constant $\theta$, i.e.

$$\int_{t_n}^{t_n+\Delta t_n} f(\lambda)d\lambda = \theta \quad (5)$$

which is exactly what the IFC does. We changed the notation from $nT$ to $t_n$ because the new representation can be aperiodic. Substituting eqn. 5 in eqn. 4 yields

$$f(t)|_{t=t_n} = \sum_{n=-\infty}^{\infty} \int_{t_n}^{t_n+\Delta t_n} f(\lambda)d\lambda \quad (6)$$

This shows that the normalization, which is strictly enforced by the hardware, has destroyed the sampling relationship ($f(t_n)$ is always a multiple of $\theta$), but notice that now *there is no need for an external clock* as long as we reset the integration when the constraint is met, and repeat the process. As a first approximation let us think that the integration up to $\theta$ creates rectangles of constant area $\theta$ with an height given by $f(t_n)$ and base $\Delta t$, such that $f(t_n) \cdot \Delta t_n = \theta$. If we measure $\Delta t_n$ precisely, we *still can find the amplitude of the continuous time signal at time $t_n$*. In order to measure time precisely, the IFC puts out a pulse when the capacitor is reset, and *the process creates a pulse train*. The differences between pulses are $\Delta t_n$ apart, and from this we can infer by normalization the amplitude $f(t_n)$. Our previous work shows that we still can recover $f(t)$ from the pulse train (eqn. 2) with an error as small as required, so this process is an alternative to Nyquist sampling [10], [11].

It is interesting that the IFC has dual properties compared with the conventional A/D because it has much simpler implementation at the frontend, but requires more computation at the backend. For portable applications it is therefore preferred both by the lower power as well as by the lower data rates for some signals. The IFC also advances signal representation theory. In signal processing, the signal decomposition in terms of delta and Heaviside functions is very well known and thoroughly studied. The IFC provides a different picture: any finite bandwidth signal can also be exactly reconstructed by rectangles of constant area, where the ratio of the amplitude and the base change with the signal shape. This intuition is going to be critical for the definition of the algebra on pulse trains.

III. ALGEBRA ON IFC PULSE TRAINS

Currently, the normal way to operate with pulse trains [25], [26] is with a neuromorphic style of computation, which implements learned input-output mappings, and assumes a stochastic model for the events. Alternatively, *we propose to operate algebraically with pulse trains, using a deterministic (functional) framework*. This is not trivial because pulse trains are functions of continuous time; therefore, this means defining an algebra of continuous linear operators (operator algebras) [27]. Besides being complicated, it will likely be computationally impractical as an alternative for arithmetic implementations. However, the pulse trains created by the IFC have a very important property that makes the distance between consecutive pulses equal to the threshold $\theta$, which simplifies greatly the definition of the operator algebra. Since operating algebraically with successive pulse time differences corresponds to operating algebraically with the amplitude values of the input signal, when the goal is to pointwise add input signals we will sequentially add pairs of pulse trains; when the goal is to pointwise multiply analog signals, we will sequentially multiply pairs of pulse trains. This has the added advantage that the computation is online, apart from a variable delay given by the longest of the pulse pair distances. The difficulty is that we need to include in the resulting pulse trains extra pulses that still comply with the area constraint, i.e. which would be produced by an IFC with the same parameters working on the sum (or product) analog signal. This outlined direction takes us to a much more familiar territory. We will need to define an algebra over IFC pulse trains in continuous time using consecutive differences between pulses, which are real numbers. If we are successful, then we have created an isomorphism between the algebra on signal amplitudes, and the pulse trains created by the IFC.

The proposed pulse domain addition relies on the fact that the time between two pulses satisfies the area constraint; therefore, to add pulse trains, it is imperative to relate pulse differences to areas to find out when to include pulses in the time line resulting from the binary operation of addition. Because pulses occurring in two signals are asynchronous, it is also necessary to quantify carryovers between subsequent evaluations. Therefore, we start with preliminary definitions that will allow the implementation of the binary operations in pulse trains.

*A. Definitions*

*Pulse train:* Let us define a pulse train P as an ordered sequence of events over time with positive (+) or negative (-) polarity

$$P = \{p_{t_k}\} \quad (7)$$

where $t_k \in \Re, p = \pm 1, k = -\infty, \ldots, 0, 1, \ldots \infty$. Apart from polarity, which can be handled separately, the time between pulses is our main interest. Notice that P is an ordered set, so we cannot change the element order. In order to enforce the time evolution, let us further denote as $D_k$ the time between two consecutive events (left to right order), i.e. $D_k = p_{t_k} - p_{t_{k-1}}, D_k \in \Re^+, k = 1,2, \dots \infty$. Along with the polarity of the corresponding pulses, the underlying set for our field, becomes the real numbers $\Re$.

*Constant area:* Every pulse interval, $D_k$ resulting from the binary pulse train operations must satisfy the condition in eqn. 1, i.e., $\int_{t_k+\tau}^{t_{k+1}} f(t)e^{\alpha(t-t_{k+1})}dt$ ($\alpha$ and $\tau$ are the leak factor and refractory period respectively), which implies that the area under the timing between the pulses should always be equal to one constant area. The number of constant areas resulting from binary operations governs the number of pulses in the output.

*Resultant sum area (RSA):* RSA is the sum of the rate of areas of the operands in a given interval. It yields the total constant areas in an interval due to addition of operands.

*Excess area:* It is the fraction of the constant area that remains after the occurrence of one or more pulses in the output due to addition. The excess area that remains is carried over to the next interval. It is given by: Net sum area $- \lfloor$Net sum area$\rfloor$.

*Net sum area (NSA):* NSA of an interval is given by the sum of excess area and RSA. It represents the total constant area after the occurrence of the last pulse in the output due to addition. The number of pulses in an interval resulting from addition is given by $\lfloor$Net sum area$\rfloor$, where $\lfloor . \rfloor$ is the floor operator.

### B. Assumptions

*Assumption 1:* We assume that the rate of area per unit time within every pulse interval $D_k$ is constant i.e., the threshold $\theta$ (constant area) is reached linearly. This approximation is accurate for periodic pulse trains (constant signals) as shown in Fig. 2(a). In case of aperiodic pulse trains, the approximation is very good for small $D_k$, while if $D_k$ is large it results in an error that is a fraction of $\theta$. The capacitor charging equation would be more accurate if needed. This assumption is necessary to find the rate of area of the operands in a given time within an interval.

*Assumption 2:* For mathematical convenience, the refractory period is assumed to be zero. This does not affect the proofs. In fact, the refractory period can be simply added to the resulting pulse timing without any error.

*Assumption 3:* The computation of areas is done at every pulse timing to respect the inherent time structure. This interval wise addition ensures that the changes in rate of areas i.e., inter pulse intervals between the pulse trains and changes in RSA are accounted for in the resultant pulse timings. For instance, consider the addition of three pulse trains namely augend, addend 1, and addend 2 pulse trains shown in Fig. 2(b). Among the pulse trains, the first pulse occurs at $t_{u_1}$; therefore the computation of areas for the three pulse trains is done during the time interval $(0, t_{u_1})$. The next pulse occurs at $t_{d_1}$ and hence the computation of areas is done during the time interval $(t_{u_1}, t_{d_1})$. For the pulse trains shown in Fig. 2(b), there are eight different pulse intervals at which the areas are computed to ensure that the variations in RSA are accounted for in the resultant pulse timings.

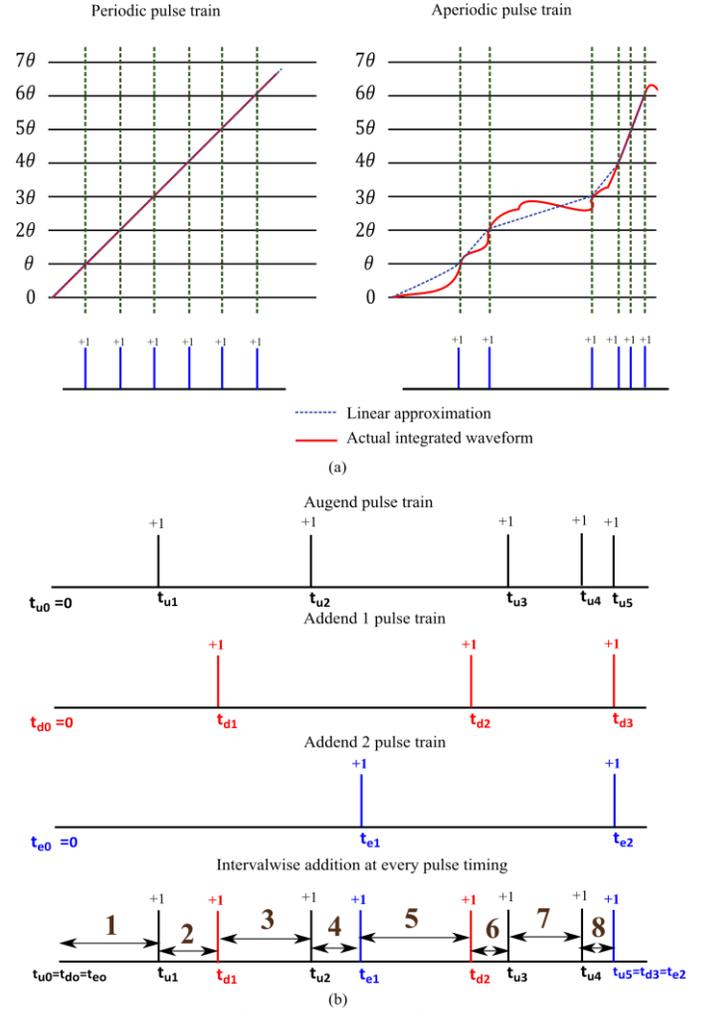

Fig. 2. (a) Illustration of the assumption of constant increase in area within every pulse interval. (b) Illustration of interval wise addition at every pulse timing.

### C. Theorem 1: Addition of pulses

For $j = 1,2,3,\cdots m$, let $t_{u_j}$ and $t_{d_1}$ denote the pulse times with positive polarities of an *augend* pulse train and *addend* pulse train respectively. Suppose that $T_k$ denotes the pulse time of the resultant sum of augend pulse train and addend pulse train, and $0 = t_{u_0}(= t_{d_0}) < t_{u_1} < t_{u_2} \cdots < t_{u_m} \leq t_{d_1}$ then

a. For $k = 1,2,3,\cdots m$, $T_k = t_{u_{k-1}} + \frac{(t_{d_1} - t_{u_{k-1}})A_k}{B_1 + A_k}$
where $A_k = (t_{u_k} - t_{u_{k-1}})$ and $B_1 = (t_{d_1} - t_{d_0})$ are the inter-pulse augend and addend durations respectively.

b. $T_{m+1} = t_{d_1}$.

Moreover, the number of pulses in the resulting sum of the two positive polarity pulse trains is $m + 1$.

*Proof:*

Consider the augend and addend pulse train shown in Fig. 3(a). In the interval $(t_{u_0}, t_{u_1})$, the rate of area per unit time due to addition is given by the sum of rate of area per unit time in the augend $(\frac{1}{A_1})$ and addend $(\frac{1}{B_1})$ respectively. The RSA in

$(t_{u_0}, t_{u_1}) = 1 + \frac{A_1}{B_1}$ is greater than one constant area. Hence, in the pulse train of the resultant sum, the first pulse occurs with positive polarity and its timing is given by

$$T_1 = \frac{A_1 B_1}{A_1 + B_1} = t_{u_0} + \frac{(t_{d_1} - t_{u_0})A_1}{A_1 + B_1} = \frac{t_{d_1} t_{u_1}}{t_{d_1} + t_{u_1}} \quad (8)$$

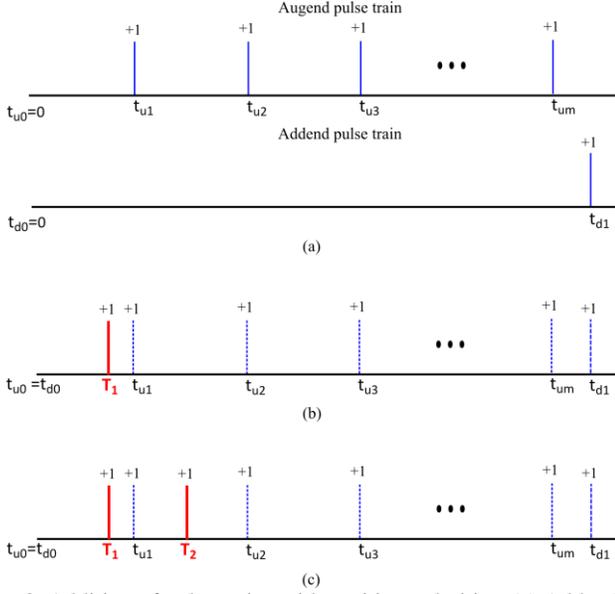

Fig. 3. Addition of pulse trains with positive polarities. (a) Addend and augend pulse trains. (b) Relative position of first pulse in resultant sum with respect to other pulse timings. (c) Relative position of second pulse in relation to other pulse timings.

The resultant pulse train at this stage is shown in Fig. 3(b) and we have

$$\text{Excess area in } (T_1, t_{u_1}) = \frac{A_1}{B_1} \quad (9)$$
$$\text{Excess area time in } (T_1, t_{u_1}) = (t_{u_1} - T_1) \quad (10)$$

In the next interval $(t_{u_1}, t_{u_2})$, the rate of area per unit time is given by $\frac{B_1 + A_2}{B_1 A_2}$ and RSA in the interval $(t_{u_1}, t_{u_2})$ is $\frac{B_1 + A_2}{B_1}$. The NSA in the interval $(T_1, t_{u_2})$ is given by the sum of excess area in $(T_1, t_{u_1})$ and the RSA in $(t_{u_1}, t_{u_2})$. From eqn. 9, we have

$$\text{NSA in } (T_1, t_{u_2}) = \frac{A_1}{B_1} + \frac{B_1 + A_2}{B_1} = 1 + \frac{\sum_{i=1}^{2} A_i}{B_1} \quad (11)$$

The NSA in $(T_1, t_{u_2})$ is greater than one constant area. Hence, the second pulse in the resultant sum occurs in $(t_{u_1}, t_{u_2})$ with positive polarity and excess area is given by

$$\text{Excess area in } (T_2, t_{u_2}) = \frac{\sum_{i=1}^{2} A_i}{B_1} \quad (12)$$
$$\text{Excess area time in } (T_2, t_{u_2}) = (t_{u_2} - T_2) \quad (13)$$

The resultant pulse train at this stage is shown in Fig. 3. The exact area required for $T_2$ can be obtained by subtracting the excess area in $(T_2, t_{u_2})$ from one constant area. From eqn. 9, we have the following relations:

$$1 - \text{excess area in } (T_1, t_{u_2}) = \frac{B_1 - A_1}{B_1} \quad (14)$$

The time taken for (1-excess area), $T_{1-ex}$ in $(T_1, t_{u_2})$ is given by the ratio of (1-excess area) in $(T_1, t_{u_2})$ to the rate of area per unit time in the interval $(t_{u_1}, t_{u_2})$. So, we have

$$T_{1-ex} = \left(\frac{B_1 - A_1}{B_1}\right)\left(\frac{B_1 A_2}{B_1 + A_2}\right) = \left(\frac{B_1 - A_1}{B_1 + A_2}\right) A_2 \quad (15)$$

The second pulse time interval $T_2 - T_1$ is given by the sum of excess area time in $(T_1, t_{u_1})$ and $T_{1-ex}$ in $(T_1, t_{u_2})$. From eqn. 10 and eqn. 15, the second pulse time $T_2$ is given by

$$T_2 = t_{u_1} + \left(\frac{t_{d_1} - t_{u_1}}{B_1 + A_2}\right) A_2 \quad (16)$$

Repeating this procedure with the interval $(t_{u_2}, t_{u_3})$, the rate of area per unit time is given by $\frac{B_1 + A_3}{B_1 A_3}$ and the RSA is $\frac{B_1 + A_3}{B_1}$. The NSA in $(T_2, t_{u_3})$ is $1 + \frac{\sum_{i=1}^{3} A_i}{B_1}$ which is greater than one constant area. Hence, the third pulse occurs in $(t_{u_2}, t_{u_3})$ with positive polarity and we have

$$\text{Excess area in } (T_3, t_{u_3}) = \frac{\sum_{i=1}^{3} A_i}{B_1} \quad (17)$$
$$T_3 = t_{u_2} + \left(\frac{t_{d_1} - t_{u_2}}{B_1 + A_3}\right) A_3 \quad (18)$$

By induction, $m^{th}$ pulse occurs with positive polarity and

$$\text{Excess area in } (T_m, t_{u_m}) = \frac{\sum_{i=1}^{m} A_i}{B_1} \quad (19)$$
$$T_m = t_{u_{m-1}} + \left(\frac{t_{d_1} - t_{u_{m-1}}}{B_1 + A_m}\right) A_m \quad (20)$$

In the interval $(T_m, t_{d_1})$ two cases arise:
*Case 1:* $t_{u_m} < t_{d_1}$
The rate of area per unit time in $(t_{u_m}, t_{d_1})$ is given by $\frac{1}{B_1}$ and the corresponding RSA is $\left(\frac{t_{d_1} - t_{u_m}}{B_1}\right)$. The NSA in $(T_m, t_{d_1})$ is $\frac{\sum_{i=1}^{m} A_i}{B_1} + \left(\frac{t_{d_1} - t_{u_m}}{B_1}\right)$ which is equal to one constant area. Hence $m + 1^{th}$ pulse occurs at $t_{d_1}$ and $T_{m+1} = t_{d_1}$.
*Case 2:* $t_{u_m} = t_{d_1}$
The NSA in $(T_m, t_{d_1})$ is $\frac{\sum_{i=1}^{m} A_i}{B_1} = \frac{t_{u_m}}{t_{d_1}} = 1$. Thus $m + 1^{th}$ pulse occurs at $t_{d_1}$ and $T_{m+1} = t_{d_1}$.
Equations 8, 16, 18, and 20 can be generalized as follows:

$$T_k = t_{u_{k-1}} + \frac{(t_{d_1} - t_{u_{k-1}}) A_k}{B_1 + A_k} \quad (21)$$
$$T_{m+1} = t_{d_1} \quad (22)$$

where $k = 1, 2, 3, \cdots m$. Thus the pulse train of the resultant sum of augend pulse train and addend pulse train has $m + 1$ pulses with positive polarities.

*Corollary 1*

Suppose that the augend pulse train has one pulse time at $t_{u_1}$ with positive polarity and the addend pulse train has $m$ pulses with positive polarities at $t_{d_i}$, $i = 1, 2, 3, \cdots m$ such that $0 = t_{u_0}(= t_{d_0}) < t_{d_1} < t_{d_2} \ldots < t_{d_m} \leq t_{u_1}$, then the resultant sum of the pulse trains has $m + 1$ pulses with positive polarities at $T_k = t_{d_{k-1}} + \frac{(t_{u_1} - t_{d_{k-1}})B_k}{A_1 + B_k}$ for $k = 1, 2, 3, \cdots m$ and $T_{m+1} = t_{u_1}$.

*Corollary 2*

Suppose that the augend pulse train has $m$ pulses with positive polarities at $t_{u_i}$, $i = 1, 2, 3, \cdots m$ and addend pulse train has $n$ pulses with positive polarities at $t_{d_i}$, $i = 1, 2, 3, \cdots n$ such that $0(= t_{d_0}) < t_{u_1} < t_{u_2} \ldots < t_{u_{m-1}} < t_{d_1} < t_{d_2} \ldots < t_{d_n} \leq t_{u_m}$, then the resultant sum of the pulse trains has $(m + n)$ pulses with positive polarities at,

a. $T_k = t_{u_{k-1}} + \frac{(t_{d_1} - t_{u_{k-1}})A_k}{B_1 + A_k}$ where $k = 1, 2, 3, \cdots m$.

b. $T_{m+k} = t_{d_{k-1}} + \frac{(t_{u_m} - t_{d_{k-1}})B_k}{A_m + B_k}$ for $k = 1, 2, 3, \cdots n - 1$.

c. $T_{m+n} = t_{u_m}$.

*Remark:* Similar results as in theorem 1, corollary 1 and corollary 2 hold for the resultant sum of two pulse trains with negative polarities.

*Corollary 3*

Suppose that the augend pulse train has $m$ pulses with negative polarities at $t_{u_i}$, $i = 1, 2, 3, \cdots m$, and addend pulse train has $n$ pulses with negative polarities at $t_{d_i}$, $i = 1, 2, 3, \cdots n$ such that $0 = t_{u_0}(= t_{d_0}) < t_{u_1} < t_{u_2} \ldots < t_{u_{m-1}} < t_{d_1} < t_{d_2} \ldots t_{d_n} \leq t_{u_m}$, then the resultant sum of the pulse trains has $(m + n)$ pulses with negative polarities.

D. *Theorem 2: Subtraction of pulse trains*

Suppose that the augend pulse train has $m$ pulses with positive polarities at $t_{u_i}$, $i = 1, 2, 3, \cdots m$ and addend pulse train has one pulse with negative polarity at $t_{d_1}$, such that $0 = t_{u_0}(= t_{d_0}) < t_{u_1} < t_{u_2} \cdots < t_{u_m} \leq t_{d_1}$ then the resultant sum of the pulse trains has $(m - 1)$ pulses with positive polarities.

*Proof.*

Consider the pulse timings and polarity of augend and addend shown in Fig. 4. In the interval $(t_{u_0}, t_{u_1})$, the rate of area per unit time due to addition is given by the sum of rate of area per unit time in the augend ($\frac{1}{A_1}$) and addend ($\frac{-1}{B_1}$) respectively. The RSA in $(t_{u_0}, t_{u_1}) = 1 - \frac{A_1}{B_1}$ is lesser than one constant area. Hence, there is no resultant pulse in $(t_{u_0}, t_{u_1})$.

$$\text{Excess area in } (t_{u_0}, t_{u_1}) = 1 - \frac{A_1}{B_1} \qquad (23)$$

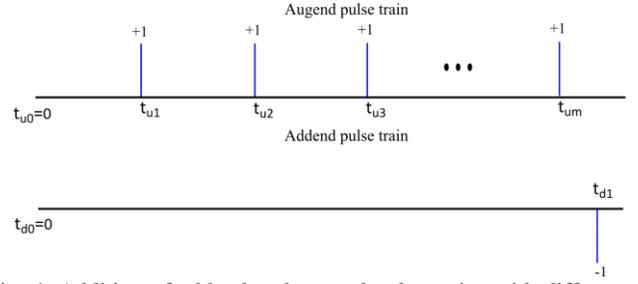

Fig. 4. Addition of addend and augend pulse trains with different polarities.

In the next interval $(t_{u_1}, t_{u_2})$, the rate of area per unit time is given by $\frac{B_1 - A_2}{B_1 A_2}$ and the corresponding RSA is $\left(1 - \frac{A_2}{B_1}\right)$. The NSA in the interval $(T_1, t_{u_2})$ is given by the sum of excess area in $(t_{u_0}, t_{u_1})$ and the RSA in $(t_{u_1}, t_{u_2})$.

$$\text{NSA in } (t_{u_0}, t_{u_2}) = 1 + \left(1 - \frac{\sum_{i=1}^{2} A_i}{B_1}\right) \qquad (24)$$

As the NSA in $(T_1, t_{u_2})$ is greater than one constant area, the first pulse occurs in $(t_{u_1}, t_{u_2})$ with positive polarity and excess area in $(T_1, t_{u_2})$ is given by $\left(1 - \frac{\sum_{i=1}^{2} A_i}{B_1}\right)$ and

$$\text{Excess area time in } (T_1, t_{u_2}) = t_{u_2} - T_1 \qquad (25)$$

The time taken for (1-excess area), $T_{1-ex}$ in $(t_{u_1}, t_{u_2})$ is given by the ratio of (1- excess area) in $(t_{u_1}, t_{u_2})$ to the rate of area per unit time in the interval $(t_{u_1}, t_{u_2})$. So, we have $T_{1-ex} = \frac{A_1 A_2}{B_1 - A_2}$. The first pulse time $T_1$ is given by the sum of excess area time in $(t_{u_0}, t_{u_1})$ and $T_{1-ex}$ in $(t_{u_1}, t_{u_2})$.

$$T_1 = t_{u_1} + \frac{A_1 A_2}{B_1 - A_2} \qquad (26)$$

In the interval $(t_{u_2}, t_{u_3})$, the rate of area per unit time is given by $\frac{B_1 - A_3}{B_1 A_3}$ and the corresponding RSA is $\left(\frac{B_1 - A_3}{B_1}\right)$. The NSA in $(t_{u_2}, t_{u_3})$ is given by $1 + \left(1 - \frac{\sum_{i=1}^{3} A_i}{B_1}\right)$ which is greater than one constant area. Hence, second pulse occurs in $(t_{u_2}, t_{u_3})$ with positive polarity at $T_2$ with excess area in $(T_2, t_{u_3})$ being given by $\left(1 - \frac{\sum_{i=1}^{3} A_i}{B_1}\right)$.

$$T_2 = t_{u_2} + \frac{(\sum_{i=1}^{2} A_i) A_3}{B_1 - A_3} \qquad (27)$$

Repeating the procedure in the interval $(t_{u_3}, t_{u_4})$, we obtain,

$$T_3 = t_{u_3} + \frac{(\sum_{i=1}^{3} A_i) A_4}{B_1 - A_4} \qquad (28)$$

$$\text{Excess area in } (T_3, t_{u_4}) = \left(1 - \frac{\sum_{i=1}^{4} A_i}{B_1}\right) \qquad (29)$$

By induction, $m - 1^{th}$ pulse occurs with positive polarity and

$$T_{m-1} = t_{u_{m-1}} + \frac{(\sum_{i=1}^{m-1} A_i) A_m}{B_1 - A_m} \quad (30)$$

Excess area in $(T_{m-1}, t_{u_m}) = \left(1 - \frac{\sum_{i=1}^m A_i}{B_1}\right)$ (31)

In the interval $(T_{m-1}, t_{d_1})$ two cases arises:

*Case 1:* $t_{u_m} < t_{d_1}$.

The rate of area per unit time in $(t_{u_m}, t_{d_1})$ is given by $\frac{-1}{t_{d_1}}$ and the corresponding RSA is $\left(-1 + \frac{t_{u_m}}{t_{d_1}}\right)$. The NSA in $(T_{m-1}, t_{d_1})$ is $\left(1 - \frac{\sum_{i=1}^m A_i}{B_1}\right) - 1 + \frac{t_{u_m}}{t_{d_1}}$ which is equal to zero. Hence, there is no pulse in $(t_{u_m}, t_{d_1})$.

*Case 2:* $t_{u_m} = t_{d_1}$.

At $t_{u_m} = t_{d_1}$, the excess area is $\left(1 - \frac{\sum_{i=1}^m A_i}{B_1}\right)$ which is less than one constant area and hence no pulse in $(T_{m-1}, t_{d_1})$.

Equations 26, 27, 28 and 30 can be generalized as follows:

$$T_k = t_{u_k} + \frac{(\sum_{i=1}^k A_i) A_{k+1}}{B_1 - A_{k+1}} \quad (32)$$

for $k = 1, 2, 3, \cdots m - 1$. Thus the resultant sum of two pulse trains, one with $m$ pulses (positive polarities) and the other with one pulse (negative polarity) yields $m - 1$ pulses with positive polarities.

*Corollary 4*

Suppose that the augend pulse train has $m$ pulses with negative polarities at $t_{u_i}$, $i = 1, 2, 3, \cdots m$ and addend pulse train has one pulse with positive polarity at $t_{d_1}$ such that $0 = t_{u_0} (= t_{d_0}) < t_{d_1} < t_{d_2} \ldots < t_{d_m} \leq t_{u_1}$ then the resultant sum of the pulse trains has $m - 1$ pulses with negative polarities.

*Corollary 5*

Suppose that the augend pulse train has $m + 1$ pulses with positive polarities at $t_{u_i}, i = 1, 2, 3, \cdots m + 1$ and addend pulse train has $n$ pulses with negative polarities at $t_{d_i}, i = 1, 2, 3, \cdots n$ such that $0 = t_{u_0} (= t_{d_0}) < t_{u_1} < t_{u_2} \ldots < t_{u_m} \leq t_{d_1} < t_{d_2} < t_{d_3} \ldots < t_{d_n} \leq t_{u_{m+1}}$ then the resultant sum of the pulse trains has $m - 1$ pulses with positive polarities and $n - 2$ pulses with negative polarities.

## IV. PROPERTIES OF ADDITION

In this section, we study the algebraic structure of pulse train addition under axioms such as identity, invertibility, commutativity and associativity.

### A. Theorem 3: Identity, inverse and commutative elements for addition

Let $G$ be the set of all pulse trains created by the IFC with $m$ pulses of positive polarities and pulse trains with $n$ pulses of negative polarities where $m, n \in \{0, 1, 2, 3, \cdots\}$, then

a. $G$ has a unique neutral pulse train $E$ in $G$ such that $P + E = E + P = P$ for all $P$ in $G$.
b. Every pulse train $P$ has a unique pulse train $Q$ in $G$ such that $P + Q = Q + P = E$.
c. For any two pulse trains $P_1$ and $P_2$ in $G$, $P_1 + P_2 = P_2 + P_1$.

*Proof.*
(i) The pulse train with no pulse and zero area is the neutral element in $G$ and denote it by $E$. Then $P + E = P = E + P$ for any $P$ in $G$.
(ii) Let $P$ be any pulse train in $G$ having $m$ pulses at $t_{u_1}, t_{u_2}, \ldots t_{u_m}$ with positive polarities and $n$ pulses at $t_{w_1}, t_{w_2}, \ldots t_{w_n}$ with negative polarities. Then the pulse train $Q$ having $m$ pulses at $t_{u_1}, t_{u_2}, \ldots t_{u_m}$ with negative polarities and $n$ pulses at $t_{w_1}, t_{w_2}, \ldots t_{w_n}$ with positive polarities is called the additive inverse of $P$ and $P + Q = Q + P = E$ and $Q$ is unique.
(iii) In computing the addition of two pulse trains $P_1, P_2$ in $G$, the pulse times of $P_1$ and $P_2$ are arranged in the ascending order. This order is same for $P_1 + P_2$ and $P_2 + P_1$. Hence the resultant sum of $P_1 + P_2$ is equal to the resultant sum of $P_2 + P_1$.

### B. Theorem 4: Associative property of addition of pulse trains

Let $G$ denote the set of all pulse trains $P_i$ with pulse times $t_i$, then the associative property holds under $G$ at every $t_i$, such that $[P_1(t_i) + P_2(t_i)] + P_3(t_i) = P_1(t_i) + [P_2(t_i) + P_3(t_i)]$.

*Proof.*

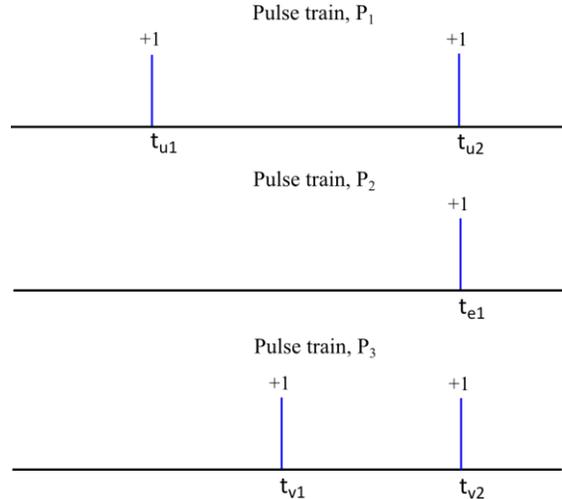

Fig. 5. Representative example of three pulse trains to demonstrate intervalwise associative property.

Consider the pulse trains shown in Fig. 5. Let $T_i$ be the timing of the pulses of $(P_1 + P_2) + P_3$ and $T_i^*$ be the timing of the pulses of $(P_2 + P_3) + P_1$. In the interval $(0, t_{u_1})$, the rate of area per unit time due to addition of $P_1$ and $P_2$ is given by $1 + \frac{t_{u_1}}{t_{e_1}}$. When the rate of area due to $P_3$ in the same interval is added to the sum of $P_1$ and $P_2$, the RSA and NSA of $(P_1 + P_2) + P_3$ in $(0, t_{u_1})$ is given by $\left[1 + \frac{t_{u_1}}{t_{e_1}}\right] + \frac{t_{u_1}}{t_{v_1}}$. This is equivalent to the RSA and NSA of $(P_2 + P_3) + P_1$ in $(0, t_{u_1})$ which is given by $\left[\frac{t_{u_1}}{t_{e_1}} + \frac{t_{u_1}}{t_{v_1}}\right] + 1$. Since the NSA in $(0, t_{u_1})$ determines the number and timing of pulses, $T_1$, the first pulse of $(P_1 + P_2) + P_3$, and $T_1^*$, the timing of the first pulse of $(P_2 + P_3) + P_1$, are same and occur with positive polarity. Similarly we can show that all the other pulse timings are same i.e., $T_i = T_i^*$ for both $(P_1 + P_2) + P_3$ and $(P_2 + P_3) + P_1$ at every pulse interval. Hence $[P_1(t_i) + P_2(t_i)] + P_3(t_i) = P_1(t_i) + [P_2(t_i) + P_3(t_i)]$

*Remark:* Let $G$ denote the set of all pulse trains $P_i$ with pulse times $t_i$. In $G$ define addition of two pulse trains as follows: Suppose $P_1(t_i)$ and $P_2(t_i)$ denote two pulse trains in the interval $(t_{i-1}, t_i)$, then $P_1(t_i) + P_2(t_i) = $ resultant sum pulse train in $(t_{i-1}, t_i)$. Under this operation $G$ forms an *Abelian group*.

interval) and then $P_3$ is added at every pulse interval to the resultant sum of $P_1 + P_2$, the pulse timings are given by $[4/3 \, ms, 120/43 \, ms, 40/9 \, ms, 56/9 \, ms, 8ms]$. Now, if $P_1$ and $P_3$ are added first and then $P_2$ is added to the resultant sum of $P_1 + P_3$, the pulse timings are given by $[4/3 \, ms, 48/17 \, ms, 76/17 \, ms, 56/9 \, ms, 8ms]$. Clearly, $(P_1 + P_2) + P_3 \neq P_1 + (P_2 + P_3)$ and hence associative property is true only for interval wise addition of all pulse trains simultaneously at every pulse timing.

## V. NUMERICAL RESULTS

In section III, we have shown that it is possible to algebraically perform addition of pulse trains in the pulse domain. However, in practice there are implementation issues that degrade the quality of addition of pulse trains. One of the pivotal questions is how to handle the continuous time asynchronous pulses. The solution adopted here uses a time stamping clock that quantizes pulse times [11]. Since addition

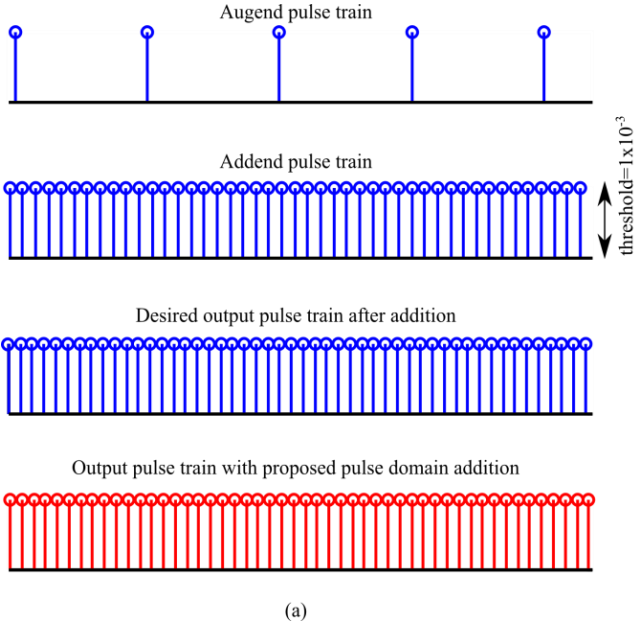 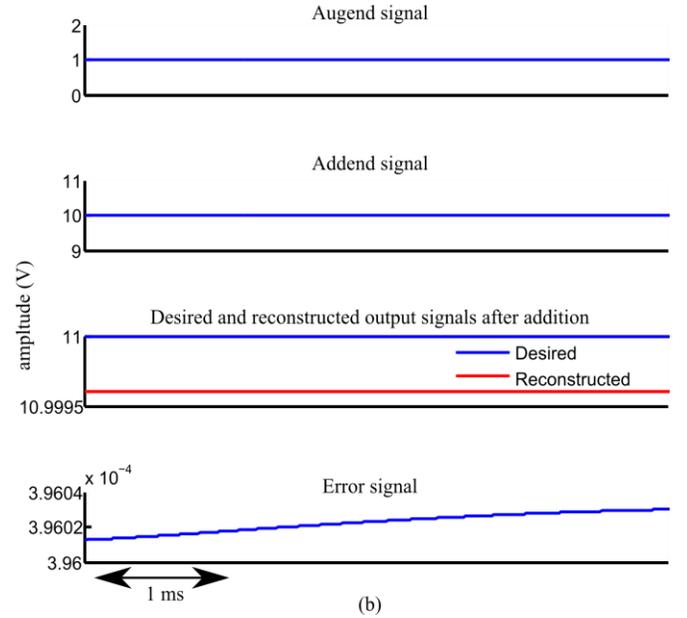

Fig. 6. Addition of periodic pulse trains. The left panel shows the addition of pulse trains corresponding to the signals in the right panel. (a) The augend, addend and desired output pulse trains are obtained with the threshold, leak factor, time stamping and refractory period set at 0.001, 40, 1µs and 0 respectively. The output pulse train calculated with the proposed pulse domain addition scheme is also shown. (b) The augend, addend and desired sum of the signals that correspond to the pulse trains in the left panel are shown in the analog domain. The reconstructed signal of the output pulse train obtained with the proposed method and the error signal are shown in the bottom of the right panel. The SNR is 88.87dB.

It is to be noted that associative property holds only for interval wise addition of all pulse trains at every pulse timing. Without assumption 3, the associative property is no longer valid. For example, let the pulse timings in Fig. 5 be given by $t_{u_1} = 8/3 \, ms, t_{v_1} = 8/2 \, ms, t_{u_2} = t_{v_2} = t_{e_1} = 8ms$. If addition is done interval wise at every pulse timing for all three pulse trains simultaneously, we have pulses at $[4/3 \, ms, 8/3 \, ms, 40/9 \, ms, 56/9 \, ms, 8ms]$ for both $[P_1(t_i) + P_2(t_i)] + P_3(t_i)$ and $P_1(t_i) + [P_2(t_i) + P_3(t_i)]$. However, if $P_1$ and $P_2$ are added first (i.e., $P_1$ and $P_2$ are added at every pulse is implemented in time, this pulse imprecision affects the timing of addend and augend pulses and results in degradation of the output pulse train, which can be studied similarly to round off noise of binary arithmetic. Unlike Nyquist ADC these errors are dependent on the input signal shape as well as on the clock frequency and, analyzing it is not straightforward as we do not have access to each sample [11]. Here we quantify the performance of pulse train addition by the signal-to-noise ratio (SNR) as the figure of merit. The SNR is given by $10 \log \frac{P_{ds}}{P_{ds} - P_{rs}}$, where $P_{ds}$ is the power of the desired signal obtained by adding the operands in the analog domain and $P_{rs}$ is the power of the reconstructed signal obtained by recovering the analog signal from the output pulse train. We also study the behavior of pulse domain addition by following an approach described in [11] and estimate the SNR under variations in the clock and threshold of the IFC.

In fig. 6, we study the addition of two periodic pulse trains that correspond to augend=1V and addend=10V. These are periodic pulse trains, as the IPIs for a given operand are the same. Also, for every one pulse in the augend, there are ten pulses in the addend. In essence by adding the two pulse trains, we are adding one and ten constant areas of the augend and addend respectively. After addition, there has to be 11 constant areas or 11 pulses for every one pulse of augend (or 10 pulses of addend) as shown in the desired output pulse train. The performance of the pulse domain addition scheme is limited by the timing imprecision in the output pulse train and the recovered signal is not exactly 11V in the analog domain resulting in an SNR of 88.87dB obtained with the threshold, leak factor, time stamping clock and refractory period set at 0.001, 40, 1μs and 0 respectively.

The variation of the SNR with the clock for a simulated IFC pulse domain addition scheme is shown in fig. 7. The input signal, threshold, leak factor and refractory period were chosen as in Fig. 6 and were kept constant to ensure a constant pulse rate. Due to limited precision of the digital arithmetic, the SNR plateaus between 1ns and 100 ns, and as expected it degrades substantially at lower clock frequencies as the exact time instant at which the threshold is reached is approximated.

the SNR degrades due to limited precision of the digital arithmetic and quantization. At higher thresholds, the number of pulses is very low and even small imprecisions in IPI results in larger errors, which also degrades the SNR. Thus, there are trade-offs in simulated IFC pulse domain scheme between error, pulse rate and choice of time stamping clocks for a given set of operands.

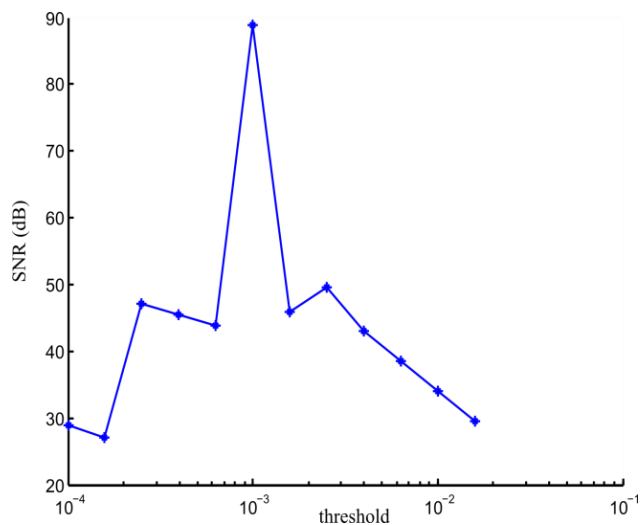

Fig. 8. Effect of threshold on SNR. The SNR was calculated for the addend and augend shown in Fig. 6 with time stamping, leak factor and refractory period set at 1μs, 40 and 0 respectively.

In fig. 9, we study the addition of two aperiodic pulse trains that correspond to sinusoidal augend and addend signals of $10\ sin(24\pi t)$ and $13\ sin(24\pi t)$. The resulting SNR with the pulse domain addition method is 42.2dB obtained with the threshold, leak factor, clock period and refractory period set at 0.001, 40, 1μs and 0 respectively. The error is relatively high at the zero crossings when compared with other regions of the signal. This is a direct consequence of assumption 1 where we assumed the threshold $\theta$ is reached linearly. Near the zero crossings, $D_k$ is relatively large; thereby, threshold is no longer linear or pseudo-linear and it results in a timing error that is a fraction of $\theta$. For instance, consider a signal transitioning from positive to negative amplitude. Just prior to crossing the zero level, the integrated area of IFC moves towards positive threshold. However, when the zero level is crossed as the signal moves into negative amplitude, the positive accumulated area of IFC has to transition into negative area to reach the negative threshold. The linear threshold assumption does not approximate this transition from positive to negative area. This results in missing pulses at zero crossings of the output pulse train as shown in fig. 9(a) and hence, the relatively higher errors near the zero level in the analog domain. Therefore if one wants high absolute accuracy in the relatively lower amplitude portions of the signal near zero level, large pulse densities will be needed, but the IFC representation loses competitiveness when compared with conventional samplers. In a positive spin, the noise floor is also most apparent in the low amplitude portions of the signal. Therefore, there is no point of representing the noise with more pulses. However, if

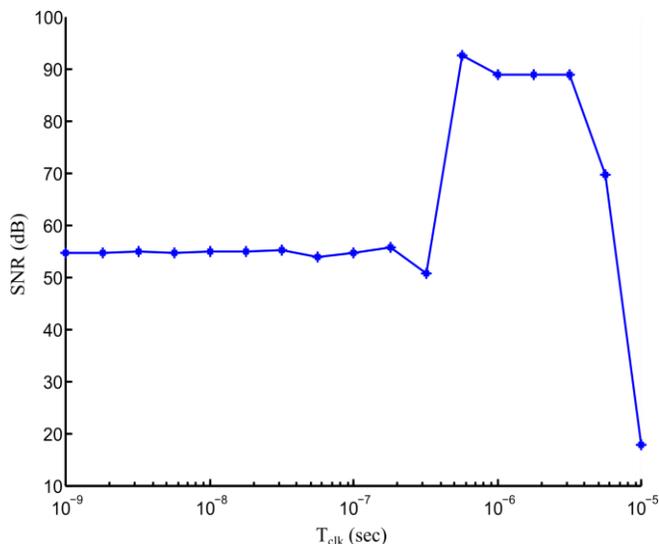

Fig. 7. Effect of time stamping on SNR. The SNR was calculated for the addend and augend shown in Fig. 6 with threshold, leak factor and refractory period set at 0.001, 40 and 0 respectively.

We also studied the variation of SNR with pulse rate by varying the threshold of the simulated IFC as shown in fig. 8. The input signal, clock period, leak factor and refractory period were chosen as in Fig. 6 and were kept constant to ensure an increasing pulse rate with decreasing threshold. At lower thresholds, the pulse rate is high and subsequently, the pulses are places closely together. As the IPI decreases towards zero, assumption 1 is substituted by the exponential capacitor charging equation, it can result in lower errors as it better approximates the transitions near the zero level. In a future follow-up paper, we will handle the finite precision studies by considering all these aspects and derive the bounds for error.

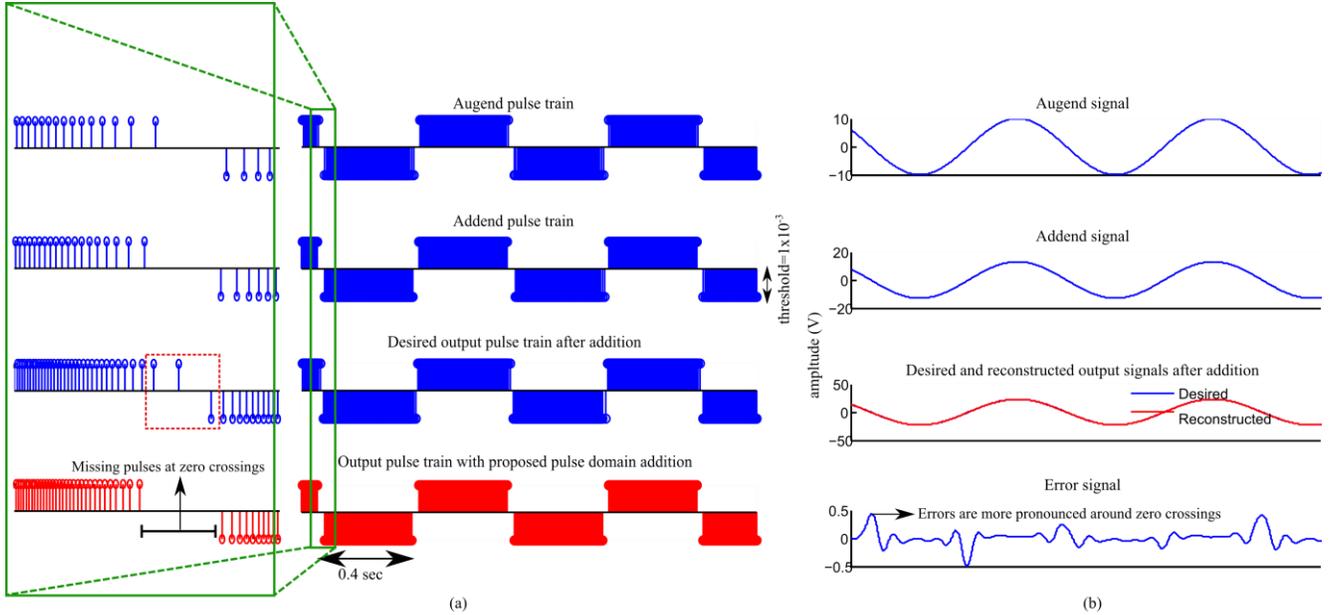

Fig. 9. Addition of aperiodic pulse trains. The left panel shows the addition of pulse trains corresponding to the signals in the right panel. (a) The augend, addend and desired output pulse trains are obtained with the threshold, leak factor, time stamping clock and refractory period set at 0.001, 40, 1μs and 0 respectively. The output pulse train calculated with the proposed pulse domain addition scheme is also shown. The zoomed version of the pulse trains shows the missing pulses at zero crossings. (b) The augend, addend and desired sum of the 12 Hz sinusoidal signals that correspond to the pulse trains in the left panel are shown in the analog domain. The reconstructed signal of the output pulse train obtained with the proposed method and the error signal are shown in the bottom of the right panel. The SNR is 42.2 dB.

Throughout the paper, we had assumed the refractory period to be zero for deriving the theorems and obtaining the results. The same results can be easily extended to non-zero refractory period. The refractory period can be ignored in the calculation of RSA and can be added back to the pulse timing of the output pulse train; thereby, the area constraint is satisfied in the output pulse train with non-zero refractory period.

## VI. Conclusion

This paper proposes a novel way of algebraically operating with +/- unitary pulse trains created by the integrate and fire converter. In this paper, we present a methodology to perform addition of pulse trains and prove that it obeys the properties of an Abelian group. Note that all the operations are defined on pairs of pulses, so it can be implemented online, with a delay given by the longest time between pulses. The addition works solely with time domain operators considering pulse polarity. The simplicity and importance of pulse train addition for signal processing come from the IFC. In fact, the IFC converts the amplitude structure of a signal into the time structure of a pulse train without loss of information, similar to the Nyquist theorem. Hence, addition of pulse trains corresponds to point wise addition of the corresponding analog signals that are fed to the pair of IFC. With the IFC constraint linking time between pulses to a constant area under the analog signal, the operator algebra can be easily implemented online, with a small delay. Therefore we effectively propose a methodology to algebraically compute with continuous amplitude and time signal sources. The advantage of algebraically operating on pulse trains is to devise alternatives to process band limited analog signals in real time and with potential ultra low power implementations without using the conventional A/D converters and digital signal processing approaches. One difficulty to achieve this goal is to develop hardware signal processing architectures to handle the continuous time pulses. Although this is not addressed in this paper, for practical implementation in hardware we propose to work locally with contiguous pulse pairs and measure time differences with counters (integers), which means that the resulting values of pulse domain addition will be rational numbers. In a future publication, we will propose hardware architectures for processing the pulse trains. We will also study its impact on the power consumption and accuracy experimentally

References

[1] E. Chris and C. H. Anderson, *Neural Engineering: Computation, Representation, and Dynamics in Neurobiological Systems*. MIT Press, 2004.
[2] N. Maghari and U.-K. Moon, "Emerging analog-to-digital converters," in *European Solid State Circuits Conference (ESSCIRC)*, 2014, pp. 43–50.

[3] J. Von Neumann, "Probabilistic logics and the synthesis of reliable organisms from unreliable components," *Autom. Stud.*, pp. 43–98, 1956.

[4] B. R. Gaines, "Stochastic Computing," in *AFIPS SJCC*, 1967, pp. 149–156.

[5] B. R. Gaines, "Techniques of Identification with the Stochastic Computer," in *Proceedings IFAC Symposium on "The Problems of Identification in Automatic Control Systems*, 1967.

[6] B. R. Gaines and J. Cleary, "Stochastic computing in neural networks Advances in Information Systems Science," *Adv. Inf. Syst. Sci.*, vol. 2, pp. 37–172, 1969.

[7] N. R. Shanbhag, R. A. Abdallah, R. Kumar, and D. L. Jones, "Stochastic Computation," in *Proceedings of the 47th Design Automation Conference*, 2010, pp. 859–864.

[8] A. Alaghi and J. Hayes, "Survey of Stochastic Computing," *ACM Trans. Embed. Comput. Syst.*, vol. 12, no. 2, 2013.

[9] W. Maass, "Liquid state machines: Motivation, theory, and applications," in *Computability in Context: Computation and Logic in the Real World*, B. Cooper and A. Sorbi, Eds. Imperial College Press, 2010, pp. 275–296.

[10] H. G. Feichtinger, J. C. Príncipe, J. L. Romero, A. Singh Alvarado, and G. A. Velasco, "Approximate reconstruction of bandlimited functions for the integrate and fire sampler," *Adv. Comput. Math.*, vol. 36, no. 1, pp. 67–78, Sep. 2011.

[11] M. Rastogi, A. Singh Alvarado, J. G. Harris, and J. C. Principe, "Integrate and fire circuit as an ADC replacement," in *2011 IEEE International Symposium of Circuits and Systems (ISCAS)*, 2011, no. 1, pp. 2421–2424.

[12] A. F. Murray, "Pulse arithmetic in VLSI neural networks," *IEEE Micro*, vol. 9, no. 6, pp. 64–74, 1989.

[13] a. Murray, "Pulse techniques in neural VLSI: a review," *[Proceedings] 1992 IEEE Int. Symp. Circuits Syst.*, vol. 5, pp. 2204–2207, 1992.

[14] M. Miura, M. Goishi, N. Chiba, and J. Shida, "A magnetic neural network utilizing universal arithmetic modules for pulse-train signal processing," *[Proceedings 1992] IEEE Int. Conf. Syst. Eng.*, pp. 499–502, 1992.

[15] A. A. Lazar, "Time Encoding with an Integrate-and-Fire Neuron with a Refractory Period," *Neurocomputing*, vol. 58–60, pp. 53–58, 2004.

[16] A. A. Lazar, "Multichannel time encoding with integrate-and-fire neurons," *Neurocomputing*, vol. 65–66, pp. 401–407, 2005.

[17] A. A. Lazar and L. T. Toth, "Perfect Recovery and Sensitivity Analysis of Time Encoded Bandlimited Signals," *IEEE Trans. Circuits Syst. Regul. Pap.*, vol. 51, no. 10, pp. 2060–2073, 2004.

[18] M. McCormick, "Digital pulse processing," MIT, 2012.

[19] D. Wei and J. G. Harris, "Signal reconstruction from spiking neuron models," in *Proceedings of the 2004 International Symposium on Circuits and Systems*, 2004, pp. 353–356.

[20] J. Harris, J. Principe, J. Sanchez, D. Chen, and C. She, "Pulse-based signal compression for implanted neural recording system," in *Proc. of IEEE International Symposium on Circuits and Systems*, 2008, pp. 344–347.

[21] A. S. Alvarado, J. C. Principe, and J. G. Harris, "Stimulus reconstruction from the biphasic integrate-and-fire sampler," in *2009 4th International IEEE/EMBS Conference on Neural Engineering*, 2009, vol. 2, no. 2, pp. 415–418.

[22] A. S. Alvarado, "Time encoded compression and classification using the integrate and fire sampler," University of Florida, 2011.

[23] G. Nallathambi and J. C. Príncipe, "Integrate and fire pulse train automaton for QRS detection.," *IEEE Trans. Biomed. Eng.*, vol. 61, no. 2, pp. 317–26, Feb. 2014.

[24] A. V. Oppenheim, A. S. Willsky, and S. Hamid, *Signals and Systems*, 2 edition. Prentice Hall, 1996.

[25] S. Mitra, S. Fusi, and I. Giacomo, "Real-time classification of complex patterns using spike-based learning in neuromorphic VLSI," *IEEE Trans. Biomed. Circuits Syst.*, vol. 3, no. 1, pp. 32–42, 2009.

[26] W. Maass, "Computing with spikes," *Special Issue on Foundations of Information Processing of TELEMATIK 8.1*, pp. 32–36, 2002.

[27] B. Blackadar, *Operator Algebras: Theory of C\*-Algebras and von Neumann Algebras*. Springer-Verlag, 2005.